\def\hatn{{\bf \hat n}}
\def\hatm{{\bf \hat m}}

\def\vecx{{\bf x}}

\def\VEV#1{{\left\langle #1 \right\rangle}}

\long\def\comment#1{}

\documentstyle[prl,aps,floats,twocolumn]{revtex}

\input{psfig.tex}
\begin{document}
\draft
\twocolumn[\hsize\textwidth\columnwidth\hsize\csname @twocolumnfalse\endcsname
\title{New Constraint on Open Cold-Dark-Matter Models}
\author{Ali Kinkhabwala\cite{aliemail}\cite{presentaddress}
and Marc Kamionkowski\cite{marcemail}}
\address{Department of Physics, Columbia University, 538 West
120th Street, New York, New York~~10027}
\date{August 1998}
\maketitle

\begin{abstract}
We calculate the large-angle cross-correlation between the
cosmic-microwave-background (CMB) temperature and the
x-ray-background (XRB) intensity expected in an open Universe
with cold dark matter (CDM) and a nearly scale-invariant
spectrum of adiabatic density perturbations.  Results are
presented as a function of the nonrelativistic-matter density
$\Omega_0$ (in units of the critical density) and the x-ray bias 
$b_x$ (evaluated at a redshift $z\simeq1$ in evolving-bias
models) for both an open Universe and a flat cosmological-constant
Universe.  Recent experimental upper limits to the amplitude of
this cross-correlation provide a new constraint to the
$\Omega_0$-$b_x$ parameter space that open-CDM models (and the
open-inflation models that produce them) must satisfy.
\end{abstract}

\pacs{PACS numbers: 98.80.Es,95.85.Nv,98.35.Ce,98.70.Vc  \hfill
CU-TP-910, CAL-666, astro-ph/9808320}
]

Determination of the matter density, $\Omega$ in units of the
critical density, has long been one of the central goals of
cosmology.  The simplest and most attractive models of inflation
\cite{inflation}, the leading paradigm for understanding the
remarkable smoothness of the Universe and the origin of its
large-scale structure, have for many years predicted the
Einstein-de Sitter value, $\Omega=1$.  However, a variety of
observations seem to suggest that nonrelativistic matter
contributes a much smaller fraction, $\Omega_0 \simeq 0.3$.  One
possible resolution is that the difference implies the existence
of a cosmological constant ($\Lambda$) that contributes a fraction
$\Omega_\Lambda\simeq0.7$ of the critical density (so that the
Universe remains flat, $\Omega=\Omega_0+\Omega_\Lambda=1$).  Another
is that some (much more complicated) models
of inflation may have produced an open Universe with
$\Omega=\Omega_0=0.3$ \cite{openinflation}.

Like ordinary inflation, open inflation produces primordial
adiabatic density
perturbations that, with the presence of cold dark matter (CDM),
give rise to the large-scale structure
observed in the Universe today.  When the amplitude of
density perturbations is normalized to the Cosmic Background
Explorer (COBE) map of the cosmic microwave background (CMB),
such ``open-CDM'' models are found to be consistent with the
amplitude and shape of the galaxy power spectrum
\cite{ks,ratra}.

In~an~Einstein-de~Sitter~Universe, large-angle CMB anisotropies
are produced by
gravitational-potential differences induced by density
perturbations at the surface of last scatter at a redshift
$z\simeq1100$ via the Sachs-Wolfe (SW) effect \cite{sw}.  In a flat
$\Lambda$ Universe \cite{kofman}, or in an open
Universe \cite{ks}, additional anisotropies are produced by
density perturbations at lower redshifts along the line of sight
via the integrated Sachs-Wolfe (ISW) effect.  Ref.~\cite{ct}
thus argued  that in a flat $\Lambda$
Universe, there should be some cross-correlation
between the CMB and a tracer of the mass distribution at low
redshifts; a similar effect also occurs in an open Universe
\cite{marc}.

The x-ray background (XRB) currently offers perhaps the best tracer of 
the mass distribution out to redshifts of a few.
Ref. \cite{bct} determined an upper limit to
the amplitude of the cross-correlation between the COBE CMB map
and the first High-Energy
Astrophysical Observatory (HEAO I) map of the 2--20 keV XRB, and
used it to constrain $\Omega_0\gtrsim0.3$ (with some
assumptions) in a flat $\Lambda$ Universe.

The ISW effect contributes a much larger fraction of
the large-angle anisotropy in an open Universe than in a flat
Universe with the same $\Omega_0$ (see Ref.~\cite{ks} and Fig.~1 in
Ref.~\cite{marc}), so one would expect
the cross-correlation to be much ($\sim5$ times for
$\Omega_0\sim0.3$) larger in an open Universe.
However, the XRB only probes the
matter distribution out to a redshift of $z\sim4$, so the
relevant quantity is the fraction of the CMB anisotropy produced
by the ISW effect at these low redshifts.  Although virtually
all of the ISW effect comes from these low redshifts in a
flat $\Lambda$ Universe, the range of redshifts over
which the ISW effect contributes in an open Universe is much
broader (cf., Fig.~2 in Ref.~\cite{marc}).  Thus, a detailed
calculation is necessary to apply the results of Ref.~\cite{bct}
to an open Universe.

In this paper, we generalize the calculation of Ref.~\cite{ct}
to an open Universe.  We present results as a function of
$\Omega_0$ and a currently uncertain bias $b_x$ of x-ray sources 
for an open and a flat $\Lambda$ Universe.  An
experimental upper limit \cite{bct} is used to constrain the
$\Omega_0$-$b_x$ parameter space for open-CDM and flat $\Lambda$CDM
models.  We show that these constraints depend only weakly on
the Hubble constant, spectral
index, uncertainties in the large-scale power spectrum, and
uncertainties in the XRB redshift distribution.  If
$\Omega_0\simeq0.3-0.4$, then x-ray sources can be no more than
weakly biased tracers of the mass distribution.  We discuss how to
apply these results to models of evolving x-ray bias.

We now detail our calculation: The fractional perturbation to
the temperature in a direction 
$\hatn$ is
\begin{eqnarray}
     \frac{\Delta T}{T}(\hatn) &=& \frac{1}{3}
     \Phi[(\eta_0-\eta_{ls})  \hatn;\eta_{ls}] +
     2 \int_{\eta_{ls}}^{\eta_{0}}
     \frac{d\Phi[(\Delta\eta)\hatn;\eta]}{d\eta}\ d\eta
     \nonumber \\
     &=&(\Delta T/T)_{SW}(\hatn) +
     (\Delta T/T )_{ISW}(\hatn),
\label{eq:deltaT}
\end{eqnarray}
where $\Delta\eta=\eta_0-\eta$, $\eta$ is the conformal time,
the subscript ``0'' denotes the value of a given parameter
today, and the subscript ``ls'' denotes the value at the surface
of last scatter at $z\simeq1100$.  In the second equality, we
have split up the temperature anisotropy into a term due to
potential perturbations at the surface of last scatter (SW) and
one due to potential fluctuations along the line of sight (ISW).
Here, $\Phi(\vecx;\eta)$ is the gravitational potential at
position $\vecx$ at conformal time $\eta$.  The 
potential is related to the density perturbation, $\delta(\vecx) 
= [\rho(\vecx) - \bar\rho]/\bar\rho$, where $\rho(\vecx)$ is the 
density at $\vecx$ and $\bar\rho$ is the mean density, through
the Poisson equation \cite{bardeen}.
Throughout, we choose the scale factor to be
$a_0=H_0^{-1}(1-\Omega_0)^{-1/2}$.  


The fractional perturbation to the XRB intensity in direction
$\hatn$ is
\begin{equation}
     \frac{\Delta X}{X}(\hatn) =\int_{\eta_{ls}}^{\eta_{0}} \, 
    g(\eta)\, \delta_x[(\Delta\eta)\hatn;\eta]\ d\eta,
\label{eq:deltaS}
\end{equation}
where $\delta_x(\vecx;\eta) = b_x \delta(\vecx;\eta)$ is the
fractional perturbation to the luminosity density of x-ray sources, and we
surmise that this is equal to some bias factor $b_x$ times the matter-density perturbation.  Here $g(\eta)$ is the selection function that determines the fraction of the XRB intensity that
comes from a conformal time $\eta$.  It is related to the XRB
redshift distribution to be discussed below.

The CMB/XRB angular auto- and cross-correlation functions are
defined by
\begin{eqnarray}
     C^{AB}(\alpha) & =& \VEV{ [\Delta A(\hatm)/ A] [\Delta B(\hatn)
     /B]}_{\hatm \cdot   \hatn=\cos\alpha} \nonumber \\
     & =& \sum_\ell {2 \ell +1 \over 4 \pi}
     C^{AB}_\ell P_\ell(\cos\alpha), 
\label{eq:correlationfns}
\end{eqnarray}
where $\{A,B\}=\{T,X\}$ are the fractional CMB/XRB intensity
perturbations, $P_\ell(\cos\alpha)$ are Legendre polynomials,
and the angle brackets denote an average over all pairs of lines
of sight $\hatm$ and $\hatn$ separated by an angle $\alpha$.
Predictions for the multipole moments are given by
\begin{equation}
     C_\ell^{AB} = {2 \over \pi} \int \, k^2 \, dk \,
     \widetilde \Theta_\ell^A(k)\, \widetilde \Theta_\ell^B(k) \, P(k),
\label{eq:powerspectra}
\end{equation}
where the CMB weight functions are \cite{ks}
\begin{eqnarray}
     \widetilde\Theta_\ell^T(k) &=&\widetilde\Theta_\ell^{SW}(k)+
     \widetilde\Theta_\ell^{ISW}(k) \nonumber \\
     &=& \left[ {3 \Omega_0 \over 
     2 (1-\Omega_0)(k^2+4)} \right]
      \Biggl[{1\over 3} \Phi_k^\ell(\eta_0-\eta_{ls}) F(\eta_{ls})
      \nonumber \\
      & & \quad +2\int_{\eta_{ls}}^{\eta_0} \, F'(\tilde\eta)
      \Phi_k^\ell(\eta_0 -\tilde\eta) \, d\tilde\eta \Biggr],
\label{eq:tildethetaT}
\end{eqnarray}
and the function $F(\eta)$ describes the time evolution of
potential perturbations; it is given in terms of the well-known 
linear-theory growth factor $D(z)$ for density perturbations
\cite{peebles} by $F(z)=(1+z)\,D(z)/D(0)$.
The XRB analog is \cite{lahav}
\begin{equation}
     \widetilde\Theta_\ell^X(k) = b_x\int_{\eta_{ls}}^{\eta_0}
     \, g(\tilde\eta) \, D(\tilde\eta) \,\Phi_k^\ell(\eta_0
     -\tilde \eta) \, d\tilde\eta,
\label{eq:tildethetaX}
\end{equation}
and for $k\gg1$ (scales smaller than the curvature scale),
$P(k) \propto k^n T^2(k)$ (with $n\simeq 1$) is the power spectrum 
for the mass distribution with $T(k)$ the transfer function
\cite{bbks}.  The functions $\Phi_k^\ell(\eta)$ are
the radial harmonics for a space of constant negative curvature
\cite{w,ks}, the curved-space analog of spherical Bessel
functions.

If the correlation functions are measured using beams with
Gaussian profiles of fwhm $\theta_{\rm fwhm}$, then factors
$W^A_\ell(\theta) W^B_\ell(\theta)$, should be
included in the sum in Eq. (\ref{eq:correlationfns}), where
$W_\ell(\theta) = \exp[-\ell(\ell+1)\sigma_b^2/2]$ is the window 
function, and $\sigma_b=0.00742 \, (\theta_{\rm fwhm}/1^\circ)$.

Since we compare the results of our calculation to the
experimental limit of Ref.~\cite{bct}, we
simply use the redshift distribution used by Ref. \cite{bct}.  This
model assumes that the universal x-ray luminosity evolves,
increasing with increasing redshift, from $z=0$
until $z_c=2.25$, and is thereafter constant up to
a maximum redshift $z_f=4$, beyond which the x-ray luminosity is
zero \cite{xray}.  The fraction of the
local x-ray flux that comes from any given differential redshift 
interval is obtained in the standard way (see, e.g.,
Ref. \cite{peebles}), and $g(\eta)$ is also obtained in the
standard way.
To assess the effects of uncertainties in the XRB redshift
distribution on the final results, we also consider two
alternative redshift distributions.  In the first,
we simply scale our canonical redshift distribution so that it
extends to a redshift $z_f=5$, instead of 4 (so the evolution
cutoff is at $z_c=2.81$).  In the second, we scale the
canonical distribution so that it extends out only to $z_f=3$
(so $z_c=1.69$).  

The power spectrum $P(k)$ is normalized so that the rms fluctuation,
$\sigma_T=\left[C^{TT}(0)\right]^{1/2}$ [calculated from
Eqs. (\ref{eq:correlationfns})-(\ref{eq:tildethetaT}) and
smoothed with a Gaussian beam with $\theta_{\rm
fwhm}=10^\circ$], matches that measured by COBE \cite{banday}.
The x-ray bias
$b_x$ must then be chosen so that the predicted rms XRB
fluctuation [calculated from Eqs. (\ref{eq:correlationfns}),
(\ref{eq:powerspectra}) and (\ref{eq:tildethetaX}) and
smoothed with $\theta_{\rm fwhm}=3.6^\circ$] matches the
empirical value.  The experimental result for the
rms XRB fluctuation from HEAO I is $\sigma_x^{\rm HEAO}=0.024$
\cite{bct}.  However, some fraction of this measured fluctuation 
amplitude must come from Poisson fluctuations in the (currently
uncertain) number density of sources that give rise to the
diffuse extragalactic XRB.  The rest is due to the large-scale
mass inhomogeneities, as traced by x-ray sources, and is what we are
interested in.  Thus, $(\sigma_x^{\rm HEAO})^2=(\sigma_x)^2 +
(\sigma_x^{\rm Poisson})^2$, so $\sigma_x \lesssim 0.024$.

To proceed, we first determine the x-ray bias that 
would be needed if all of the measured XRB fluctuation amplitude 
were due to density perturbations (i.e., if we assumed
$\sigma_x=\sigma_x^{\rm HEAO}$), and then calculate the
zero-lag cross-correlation amplitude $C^{XT}(0)$.  If some
fraction of the fluctuation amplitude is due to Poisson 
fluctuations, then the fluctuation due to density perturbations
must be smaller.  The x-ray bias required to explain the observations
must therefore also be accordingly smaller.  Since 
it is proportional to the x-ray bias, the predicted
cross-correlation amplitude must also be smaller by the same
factor.


\begin{figure}[t]
\centerline{\psfig{file=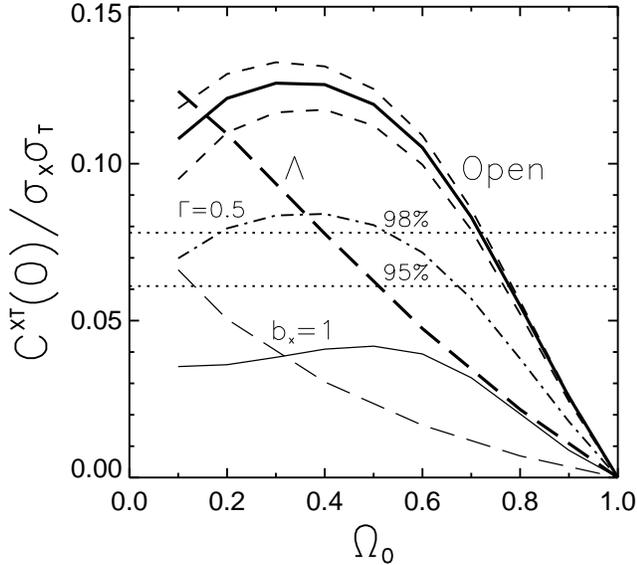,width=3.5in}}
\caption{Results for the scaled zero-lag CMB/XRB cross-correlation
     amplitude as a function of $\Omega_0$ for an open (heavy
     solid curve) and flat 
     $\Lambda$ (heavy dashed curve) Universe, both with
     $\Gamma=0.25$, assuming a constant x-ray bias and that the
     x-ray fluctuations are due
     entirely to perturbations in the density of x-ray sources.
     The upper (lower) lighter short-dash curve shows
     the result for an open Universe obtained with an
     alternative redshift distribution that extends to larger
     (smaller) redshifts.  The dot-dash curve shows the
     result for an open Universe with $\Gamma=0.5$.  The lower
     lighter solid and long-dash curves are the 
     results for an open and flat Universe, respectively,
     assuming that $b_x=1$.  The two dotted lines are 95\% and 98\%
     confidence-level upper limits from
     Ref. \protect\cite{bct}.}
\vskip -0.5cm
\label{fig:crosscorr}
\end{figure}

Fig.~\ref{fig:crosscorr} shows our results for the scaled
zero-lag CMB/XRB cross-correlation amplitude.  (See the Figure
caption for a description of the curves.)  We have checked that
our flat-Universe calculation agrees with that of Ref. \cite{bct}.  
The decrease at small $\Omega_0$ for an open Universe) occurs because of the
additional ISW contributions to $\sigma_T$ from redshifts
$z\gtrsim4$.
The predictions obtained using the alternative higher- and
lower-$z$  redshift distributions indicate that even
fairly dramatic changes to the XRB redshift distribution have
little ($\lesssim10\%$) effect on the predicted
cross-correlation amplitude.

\begin{figure}[t]
\centerline{\psfig{file=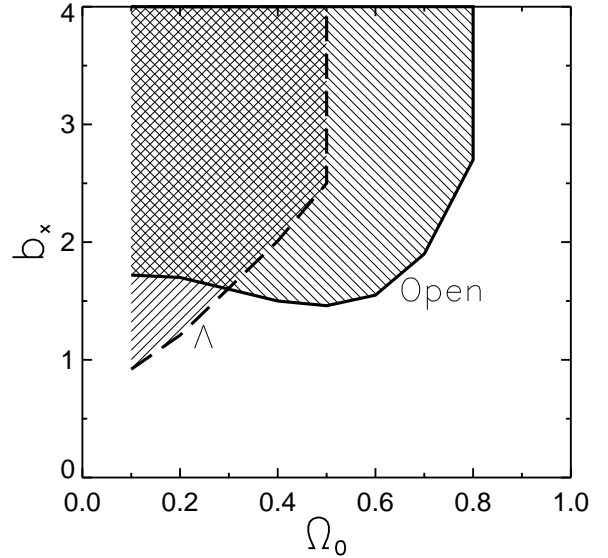,width=3.5in}}
\caption{Constraints to the $\Omega_0$-$b_x$ parameter space for 
     both an open Universe and a flat $\Lambda$ Universe.  The
     shaded regions are ruled out at the 95\% CL.  In
     evolving-bias models, the $b_x$ shown here is the value of
     the x-ray bias at $z\simeq1$.}
\vskip -0.5cm
\label{fig:parameterspace}
\end{figure}

Fig.~\ref{fig:crosscorr} shows results for two extreme
assumptions about the contribution of Poisson fluctuations, or
equivalently, the x-ray bias.  The heavy curves show predictions 
obtained by assuming that all of the measured XRB fluctuation
amplitude is due to density perturbations (i.e., no Poisson
fluctuations), which implies the largest possible bias (and
cross-correlation).  The lighter curves show results obtained if
we assume the x-ray bias is $b_x=1$; these curves provide a
lower limit to the cross-correlation amplitude as long as x-ray
sources are not antibiased.  Since the scaled
cross-correlation amplitude is proportional to the bias, the
x-ray bias inferred if we assume no Poisson fluctuations can be
obtained by taking the ratio of the maximal prediction to the
$b_x=1$ prediction.
Similarly, we can obtain the predicted cross-correlation
amplitude for any x-ray bias by interpolating between the
$b_x=1$ and maximal predictions.  Doing so, we obtain
constraints to the $\Omega_0$-$b_x$ parameter space shown in
Fig.~\ref{fig:parameterspace}, which illustrates our central
results.

We now detail how our result depends on certain model parameters.
We used a flat $n=1$ primordial power spectrum.  A smaller value
of $n$ will increase the cross-correlation amplitude relative to 
$\sigma_T$, but it will also increase $\sigma_x$.  Numerically,
if $n$ is decreased to 0.7, the scaled cross-correlation
amplitude increases by 15\% for $\Omega_0=0.4$, and conversely
for larger $n$.  

Although the generalization of a power-law primordial power
spectrum to an open Universe is not well-defined for low $k$
\cite{ks}, this uncertainty only affects the lowest CMB
multipole moments.  The effect on the XRB is small because the
spectrum of density fluctuations leans much more to smaller
scales (because of the Poisson equation) than that for the
potential perturbations that give rise to CMB anisotropies.
Numerical calculations show that the curves in
Fig.~\ref{fig:crosscorr} are changed by only a 
few percent (for $\Omega_0\simeq0.4$) if alternative low-$k$ 
power spectra from open-inflation models are used.  
If, however, a fraction $f$ of the CMB variance is due to
gravitational waves, then the lower bounds in
Fig.~\ref{fig:parameterspace} are increased by a factor
$(1-f)^{-1}$.

There is some ambiguity concerning the smallest
redshift at which the diffuse XRB begins (since nearby sources
are subtracted).  In our calculations, we assumed that the XRB
distribution extends all the way down to $z=0$.  If the 
theoretical predictions are repeated assuming the XRB
distribution vanishes below $z=0.1$, the predicted
cross-correlation amplitude would increase quite significantly,
but the x-ray bias would also increase by roughly the same
amount.  Thus, this uncertainty would similarly have no
significant effect on the results shown in
Fig.~\ref{fig:parameterspace}.

How do the results depend on the parameter $\Gamma=\Omega_0 h$
that determines the peak of the present-epoch power spectrum?
If we take a very conservative upper limit 
of $h\lesssim1$, then $\Gamma\lesssim0.5$ for $\Omega_0\lesssim 
0.5$.  The dot-dash curve in Fig.~\ref{fig:crosscorr}
shows the result for $\Gamma=0.5$.  Moreover, the
reduction in the cross-correlation from the $\Gamma=0.25$ curve
can be attributed almost entirely to the reduction in the x-ray
bias.  Thus, Fig.~\ref{fig:parameterspace} will look roughly the 
same for any other reasonable value of $\Gamma$.  Since the
baryon density has only a weak effect on the power spectrum, the
results are similarly independent of the baryon density.

So far, we have taken the x-ray bias to be constant, but it has
been suggested that the x-ray bias is evolving
\cite{biasevolution,lahav}.  We have repeated our calculations
for a variety of evolving-bias models.  In each case, our
results can be reproduced by identifying our $b_x$ with the
value of the x-ray bias at a redshift $z\simeq1$ in
evolving-bias models.

To conclude, we have carried out the first calculation of the
amplitude of the CMB/XRB cross-correlation function in open-CDM
models and used an experimental upper limit to place new
constraints to the $\Omega_0$-$b_x$ parameter space in both open
CDM and $\Lambda$CDM models.  In models with evolving x-ray
bias, our $b_x$ is the bias at a redshift $z\simeq1$.  
We have shown that the excluded
regions of this parameter space are no more than weakly affected 
by uncertainties in the XRB redshift distribution, power
spectrum, Hubble constant, or baryon density.
Fig. \ref{fig:parameterspace} shows that if
$\Omega_0\simeq0.3$, then the sources that give rise to the XRB
can be no more than weakly biased tracers of the mass
distribution (unless there is a significant gravitational-wave
background).  If the high-redshift AGN that give rise to the
XRB have biases $b_x\sim3$ like other high-redshift populations
like clusters, radio sources \cite{cress}, or LBG galaxies
\cite{steidel}, then low-density CDM models will be in trouble.

The limiting factor in providing a model-independent constraint
to $\Omega_0$ is currently the uncertain Poisson contribution to 
the the XRB fluctuation amplitude, or equivalently, the
uncertain x-ray bias.  Ideally, one would remove
this uncertainty by identifying all of the sources that
contribute to the diffuse extragalactic XRB.  Fortunately, data
from forthcoming satellite experiments should help make progress 
toward this goal.

\medskip
We thank S. Dodelson, D. Helfand, and D. Spergel for useful
comments.  A.K. was supported by the Columbia Rabi Scholars
Program which is funded by the Kann Rasmussen Foundation.
M.K. was supported by a U. S. DoE OJI Award, DE-FG02-92ER40699,
NASA grant NAG5-3091, and the Alfred P. Sloan Foundation.

\end{document}